\documentclass[twocolumn,aps,prapplied,showpacs]{revtex4-1}

\usepackage{amssymb}
\usepackage{mathrsfs}
\usepackage{lipsum}
\usepackage{graphicx}
\usepackage[caption=false]{subfig}
\usepackage{mathtools}
\usepackage{epstopdf}
\usepackage{xcolor}
\usepackage[colorlinks,urlcolor=blue]{hyperref}
\DeclareGraphicsExtensions{.pdf,.jpg,.png,.eps}
\usepackage[toc,page,title,titletoc,header]{appendix}

\begin{document}

\title{Optimal binary strategy for angular displacement estimation based upon fidelity appraisal}

\author{Jian-Dong Zhang}
\affiliation{Department of Physics, Harbin Institute of Technology, Harbin, 150001, China}
\author{Zi-Jing Zhang}
\email[]{zhangzijing@hit.edu.cn}
\affiliation{Department of Physics, Harbin Institute of Technology, Harbin, 150001, China}
\author{Long-Zhu Cen}
\affiliation{Department of Physics, Harbin Institute of Technology, Harbin, 150001, China}
\author{Jun-Yan Hu}
\affiliation{Department of Physics, Harbin Institute of Technology, Harbin, 150001, China}
\author{Yuan Zhao}
\email[]{zhaoyuan@hit.edu.cn}
\affiliation{Department of Physics, Harbin Institute of Technology, Harbin, 150001, China}

\date{\today}

\begin{abstract}
We report on an optimal binary strategy for angular displacement estimation.
The measuring system is a modified Mach-Zehnder interferometer fed by a coherent state carrying orbital angular momentum, and two Dove prisms are embedded in two paths. 
In terms of the fidelity appraisal, parity detection and Z detection are discussed, and the optimal estimation strategy is presented.
We additionally study the effects of several realistic scenarios on the fidelity, including transmission loss, detection efficiency, dark counts, and those which are a combination thereof.
Finally, we exhibit a proof of principle experiment and perform Bayesian estimation on data processing. 
The experimental results imply resolved enhancement by a factor of 1.86 suggesting super-resolving signal. 
Meanwhile, on the trial greater than 500, we show that the angular displacement can be precisely estimated via acquired outcome information.  

\end{abstract}


\maketitle

\section{Introduction}

Orbital angular momentum (OAM) of light \cite{PhysRevA.45.8185} is of great academic  significance, and paves the way to a number of applications, optical communication \cite{PhysRevLett.89.240401, PhysRevA.88.032305, Willner:15}, computational imaging \cite{PhysRevLett.110.043601, Hell:94}, particle manipulation \cite{PhysRevLett.97.170406}, quantum simulation \cite{PhysRevA.75.052310, PhysRevLett.110.263602, PhysRevA.81.052322}, to name a few.
Almost all of the above applications stem from an exotic feature of OAM, which has a limitless number of orthogonal states associated with the integer values of $\ell$, also known as quantum number or topological charge. 
This provides OAM with an advantage over spin angular momentum (SAM), which only has two independent states: spin up and spin down.
Photons that are eigenstates of OAM, originate as a result of the spatial wavefront distribution.
The macroscopic embodiments are Laguerre-Gaussian beam, Bessel-Gaussian beam, and other light beams so long as the beams themselves have the spiral phase term $\exp \left( {i\ell\varphi } \right)$ with azimuthal angle $\varphi$.
To such a phase structure there corresponds to a central dark nucleus in the OAM beam, and each photon carries angular momentum $\ell\hbar $, which can be used to encode more information into a single photon.

Over the past few years, OAM-based quantum metrology has also been extensively studied, especially for angular displacement estimation \cite{lavery2013detection, lavery2014observation,PhysRevLett.112.200401,
	d2013photonic, zhang2017optimal}.
Theoretically, the Heisenberg limit can be achieved by several exotic quantum states carrying OAM, e.g., two-mode squeezed vacuum, entangled coherent, and N00N states.
However, from the view of sensitivity, many researches have illustrated that a high-power coherent state downplays the quantum states as it is very difficult to produce a quantum state with large photon number \cite{escher2011quantum, lang2013optimal}.
Hence, some super-resolution protocols using an OAM coherent state and binary detection strategies have been proposed \cite{zhang2017optimal}.
Under the circumstances, the character of super-resolution is well demonstrated, while quantum detection strategy implements a single-fold super-resolution peak, and OAM generalizes this feature in accordance with double quantum number $\ell$, i.e., a $2\ell$-fold super-resolution peak is realized.
On the other hand, the sensitivities of the two binary detection strategies$\---$Z detection  \cite{cohen2014super} and parity detection \cite{bollinger1996optimal, gerry2010parity}$\---$are saturated by the shot noise limit, so that we can not determine the pros and cons of the two strategies from the view of sensitivity. 
Moreover, some researches show that the standard deviation $\delta \theta$ may not be an adequate criterion to characterize the phase sensitivity in an interferometer when multiple peaks are presented in the phase probability distribution \cite{pezze2006phase, pezze2005mach, yurke19862, kim1998influence}, and OAM coherent state along with binary detection strategy clearly belongs to this category.
Therefore, we try to appraise the two strategies by replacing standard deviation with fidelity estimation.

\begin{figure*}[htbp]
	\centering
	\includegraphics[width=0.9\textwidth]{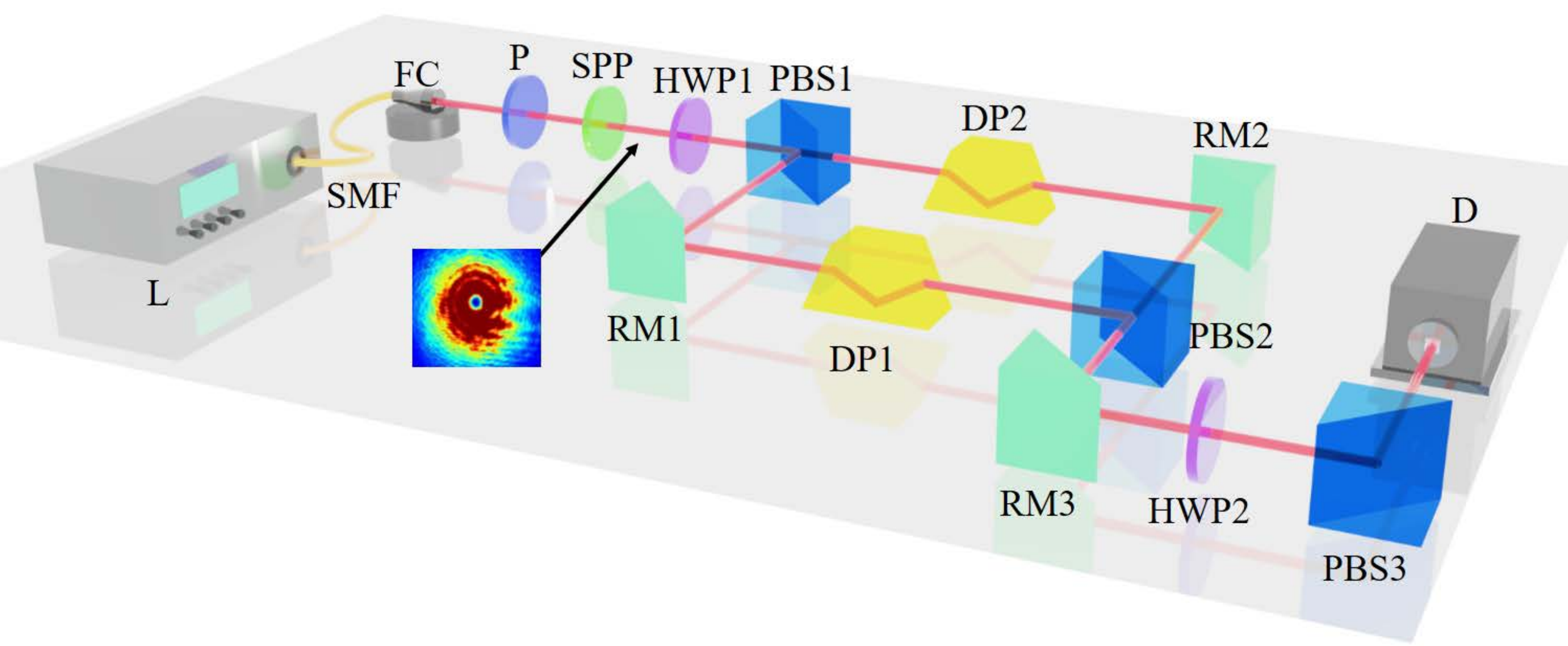}
	\caption{Schematic of the angular displacement estimation protocol. The subgraph is the intensity distribution of the modulated OAM beam obtained by a CCD camera. The devices are abbreviated as: L, laser; SMF, single mode fiber; P, polarizer; FC, fiber coupler; SPP, spiral phase plate; HWP, half wave plate; PBS, polarizing beam splitter; DP, Dove prism; RM, reflection mirror; D, detector.}
	\label{f1}
\end{figure*}

In this paper, we report using fidelity \cite{bahder2006fidelity, simon2008dispersion, bahder2011phase} (the Shannon mutual information between the angular displacement $\theta$ and the measuring outcomes) to appraise two binary detection strategies whose sensibilities are on a par with each other under the standard deviation metric. 
A simple understanding for the mutual information is the amount of sender's original information which can be obtained by receiver from the acquired distortion information. 
It follows that high fidelity is closely related to excellent parameter sensitivity, and we rely on the size of the fidelity to determine a more suitable strategy from the two strategies.

The remainder of this paper is organized as follows: in Sec. \ref{s2}, we briefly introduce OAM-enhanced angular displacement estimation protocol and two binary detection strategies.
The comparison between the two strategies is exhibited in Sec. \ref{s3}, and the optimal one is presented.
In Sec. \ref{s4} we analyze the impacts of several realistic scenarios on the optimal strategy, including transmission loss, detection efficiency, dark counts, and those which are a combination thereof.
In Sec. \ref{s5}, we demonstrate a proof of principle experiment and perform Bayesian estimation on data processing, the influence factors in experiment are also discussed. 
Finally, we summarize our work with a concise conclusion in Sec. \ref{s6}.

\section{Fundamental principle}
\label{s2}

\subsection{Estimation protocol and detection strategy}

Consider a modified Mach-Zehnder interferometer whose input is a coherent state carrying OAM and two Dove prisms in its arms, as illustrated in Fig. \ref{f1}. 
A light beam generated from the laser is coupled into a single mode fiber, the output is collimated, and then it is sent to a polarizer.
The single mode fiber and the polarizer are responsible for purifying the beam's space mode and polarization, respectively.
OAM degree of freedom of the beam is added via a spiral phase plate, in turn, the input state can be written as ${\left| {{\alpha _\ell}} \right\rangle _A}{\left| 0 \right\rangle _B}$ for quantum number $\ell$.
Subsequently, the beam polarization is changed to diagonal direction with a half wave plate, and then it enters a polarizing Mach-Zehnder interferometer, where the relative angular displacement $\theta$ between the two prisms is the parameter we would like to estimate.
For such an evolution process, the prism in mode $A$ is rotated with displacement $\theta$ and that in mode $B$ is immobile, the operator corresponds to the following form: $\hat U = \exp \left( {i2\ell{{\hat a}^\dag }\hat a\theta } \right)$ \cite{PhysRevA.96.052118}. 
After the interferometer, the second half wave plate rotates the beam's polarization from horizontal (vertical) direction to diagonal (anti-diagonal) one.
Then, the two polarized modes interfere with each other due to the presence of third polarizing beam splitter, either of the two outputs can be detected for estimation.
In terms of the above analysis, the output state reads $\left| \psi  \right\rangle  = {\left| {i{\alpha _\ell}\cos \left( {\ell\theta } \right)} \right\rangle _A}{\left| {i{\alpha _\ell}\sin \left( {\ell\theta } \right)} \right\rangle _B}$, further, this state has the following expression in the Fock state representation,

\begin{equation}
\left| \psi  \right\rangle  = {e^{ - \frac{N}{2}}}\sum\limits_{x,y = 0}^\infty  {\frac{{{{\left[ {i{\alpha _\ell}\cos \left( {\ell\theta } \right)} \right]}^x}{{\left[ {i{\alpha _\ell}\sin \left( {\ell\theta } \right)} \right]}^y}}}{{\sqrt {x!{\kern 1pt}y!} }}} \left| {x,{\kern 1pt} y} \right\rangle. 
\label{1}
\end{equation}
Where $N = {\left| {{\alpha _\ell}} \right|^2}$ is mean photon number of the OAM coherent state, and two-mode tensor product $\left| {x,{\kern 1pt}  y} \right\rangle  \equiv  {\left| x \right\rangle _A}\otimes{\left| y \right\rangle _B}$. 

Z detection and parity detection are two excellent binary strategies, neither the former nor the latter cares about exact photon number. 
Z detection only focuses on whether there are photons arriving, while parity detection  merely pays attention to the parity of measured photon number. 
Take port $B$ as an instance, the operators for Z detection and parity detection are ${\rm{\hat Z}} = {\left| 0 \right\rangle _{BB}}\left\langle 0 \right|$ and $\hat \Pi  = \exp \left( { - i{\pi }{{\hat b}^\dag }\hat b} \right)$, respectively.
The probability of count outcome is obtained from the formula $P = \left\langle \psi  \right|\hat O\left| \psi  \right\rangle $ with a projective operator $\hat O$.
Therefore, zero and nonzero counts in Z detection are given by
\begin{eqnarray}
{p}\left( {\textrm{zero}|\theta } \right) &&= \exp \left[ { - N{{\sin }^2}\left( {\ell\theta } \right)} \right],
\label{2}
\\ 
{p}\left( {\textrm{nonzero}|\theta } \right) &&= 1 - \exp \left[ { - N{{\sin }^2}\left( {\ell\theta } \right)} \right].
\label{3}
\end{eqnarray}

Similarly, we can calculate the probabilities of even counts and odd ones in parity detection,
\begin{eqnarray}
{p}\left( {\textrm{even}|\theta } \right) &&= \frac{1}{2}\left\{ {1 + \exp \left[ { - 2N{{\sin }^2}\left( {\ell\theta } \right)} \right]} \right\},  
\label{4}
\\ 
{p}\left( {\textrm{odd}|\theta } \right) &&= \frac{1}{2}\left\{ {1 - \exp \left[ { - 2N{{\sin }^2}\left( {\ell\theta } \right)} \right]} \right\}. 
\label{5}
\end{eqnarray}

\subsection{Fidelity of detection strategy}

\begin{figure}[htbp]
\centering
\includegraphics[width=8cm]{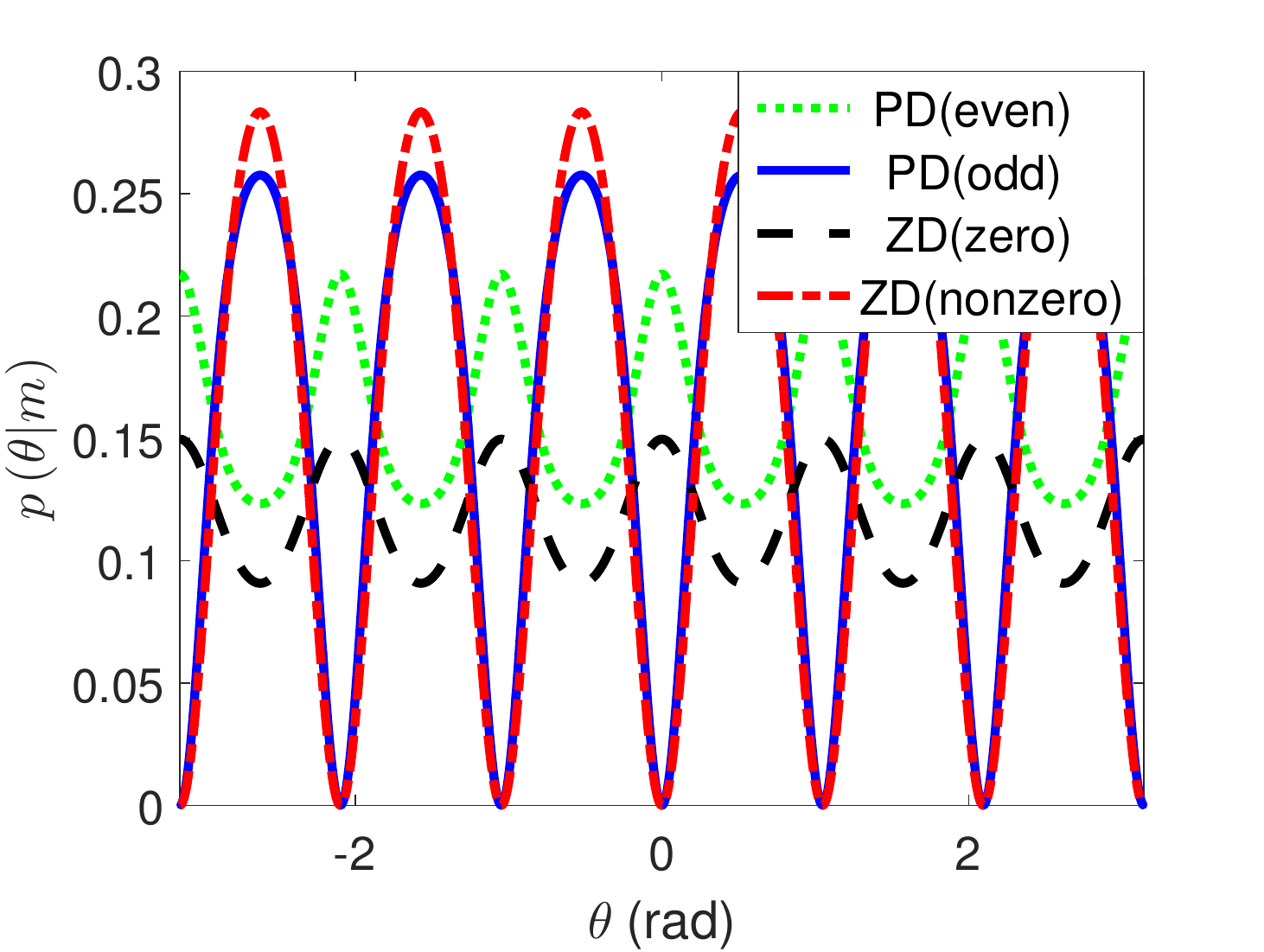}
\caption{Conditional probability density $p\left( {\theta |m} \right)$ as a function of angular displacement $\theta$ in the case of $N=1$ and $\ell=3$. Here we assume that there is no priori information. PD, parity detection; ZD, Z detection.}
\label{f2}
\end{figure}

Now we turn to the fidelity calculation, we start off with simply validating the multiple-peak character of the two strategies.
According to Bayes' rule, the conditional probability density $p\left( {\theta |m} \right)$ toward angular displacement $\theta$  and the outcome $m$ is given by \cite{bahder2006fidelity} 
\begin{equation}
p\left( {\theta |m} \right) = \frac{{p\left( {m|\theta } \right)p\left( \theta  \right)}}{{\int_{ - \pi }^\pi  {d\theta 'p\left( {m|\theta '} \right)p\left( {\theta '} \right)} }},
\label{6}
\end{equation}
where ${p\left( \theta  \right)}$ denotes a priori information of probability density for angular displacement $\theta$, over the interval $ - \pi  \le \theta  \le \pi $. 
In this paper, we assume that there is no priori information about angular displacement, i.e., $p\left( \theta  \right) = {1 \mathord{\left/{\vphantom {1 {2\pi }}} \right.\kern-\nulldelimiterspace} {2\pi }}$. 
By substituting the probabilities of Eqs. (\ref{2})$\--$(\ref{5}) into the probability density ${p\left( {m|\theta } \right)}$ of Eq. (\ref{6}), we obtain the conditional probability densities for the two strategies. 
Figure \ref{f2} displays that, in one period, the number of multi-peak is equal to double quantum number $\ell$.

\section{Strategy comparison}
\label{s3}

\begin{figure}[htbp]
\centering
\includegraphics[width=8cm]{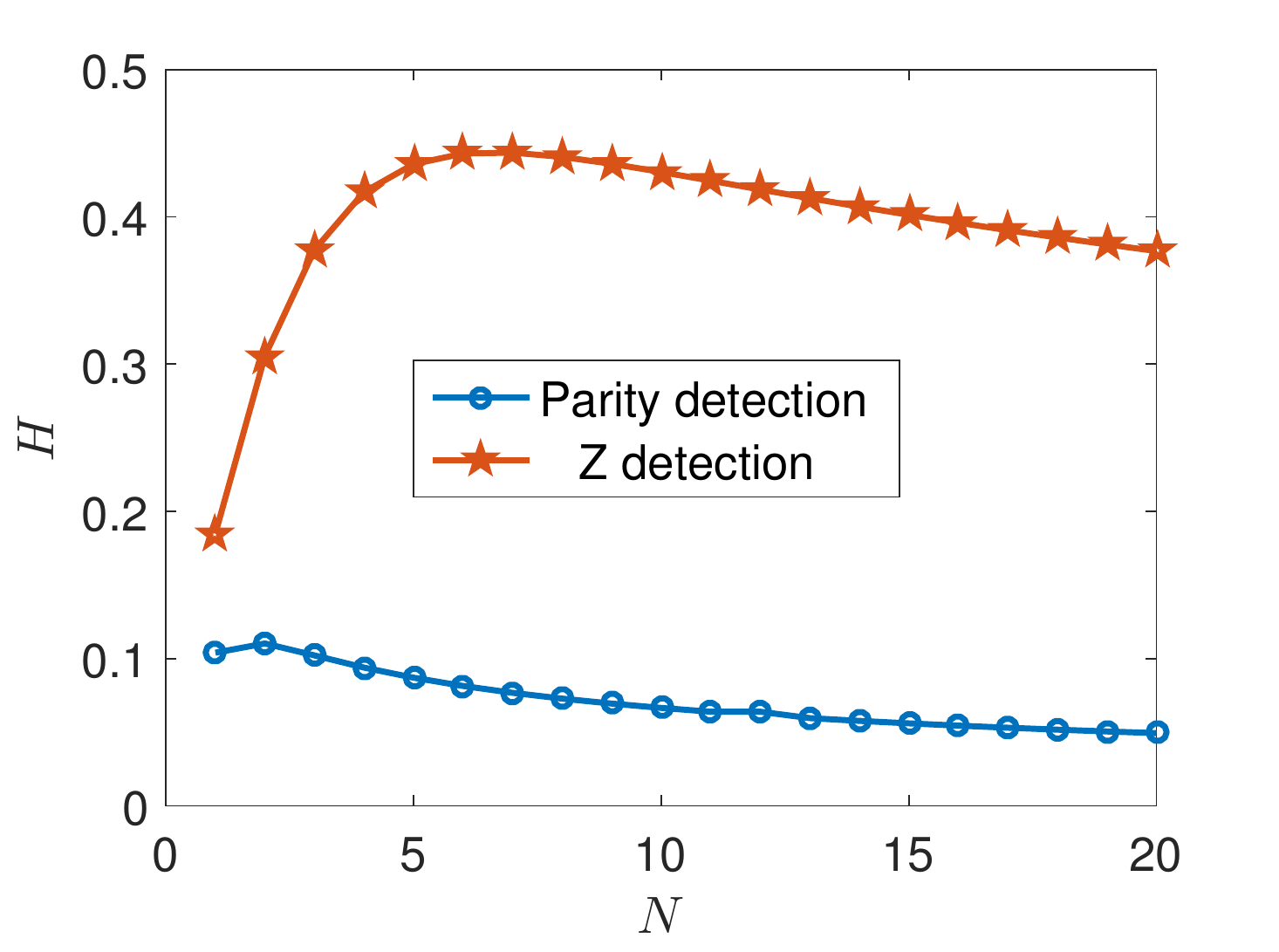}
\caption{The fidelity $H$ as a function of the mean photon number $N$, where $N$ ranges from 1 to 20 and $\ell$ is an arbitrary integer. There is also no priori information.}
\label{f3}
\end{figure}

In this section, we compare the two strategies and ascertain the optimal one by calculating fidelity. 
To begin with we give the definition of the fidelity \cite{bahder2006fidelity}, also called Shannon mutual information,
\begin{equation}
H = \sum\limits_m {\int_{ - \pi }^\pi  {d\theta p\left( {m|\theta } \right)p\left( \theta  \right){{\log }_2}\left[ {\frac{{p\left( {m|\theta } \right)}}{{\int_{ - \pi }^\pi  {d\theta 'p\left( {m|\theta '} \right)p\left( {\theta '} \right)} }}} \right]} }. 
\label{7}
\end{equation}
In views of the definition, Eqs. (\ref{2})$\--$(\ref{5}), and the assumption about the priori information, we can calculate the fidelities of the two strategies.
Figure \ref{f3} presents the fidelity variation with an increasing mean photon number $N$. 
One can find that the fidelity of Z detection is superior to that of parity detection, i.e., Z detection supplies more information about the angular displacement than parity detection from the measuring outcomes.
Meanwhile, the fidelity is independent of $\theta$, in that it is a mean value over both all possible angular displacement $\theta$ and all probabilities of measuring outcomes toward a given detection strategy.
We also confirm that this conclusion is still applicable when the mean photon number is large by simulation, e.g., the fidelity of Z detection sits at $1.053 \times {10^{ - 1}}$ and that of parity detection sits at $ 7.68 \times {10^{ - 3}}$ in the case of $N=1000$.

Overall, we can infer that Z detection is better than parity detection via analyzing parameter sensitivity from the perspective of the fidelity. 
This is a conclusion to be congratulated, at least for the realization of realistic detection,  
as parity detection generally requires the use of a photon-number-resolving detector \cite{achilles2004photon, liu2017fisher, zhang17effects}, whereas Z detection can be achieved only with an avalanche photodiode in Geiger mode (Gm-APD), a kind of detector that only presents whether the photons arrive.

\section{The effects of realistic factors}
\label{s4}

It is inevitable that an estimation system in practical application is affected by various realistic scenarios \cite{Gard2017}. 
In foregoing section, Z detection has been proved to be the optimal strategy.   
Here we study the effects of several scenarios on the fidelity, transmission loss, detection efficiency, dark counts, and those which are a combination thereof$\---$as the most commonly realistic scenarios$\---$are taken into account in this section.

\subsection{Transmission loss}

Transmission loss exists in all kinds of interferometric precision measurements. 
We can model this scenario by inserting two fictitious beam splitters of transmissivities $T_A$ and $T_B$ in each mode before the second polarizing beam splitter \cite{feng2014quantum, kacprowicz2010experimental},
$L_A=1-T_A$ and $L_B=1-T_B$ are the lossy ratios of two paths.
In turn, the output of port $B$ is given by
\begin{equation}
{\left| \psi  \right\rangle _B} = \left| {{{i{\alpha _\ell}}}\left( {\sqrt {{T_A}} {e^{ - i\ell\theta }} - \sqrt {{T_B}} {e^{i\ell\theta }}} \right)}/{2} \right\rangle,
\label{8}
\end{equation}
and the probability of detecting zero count is
\begin{equation}
{p_1}\left( {\textrm{zero}{\rm{|}}\theta } \right) = \exp \left( { - \frac{1}{4}{{\left| {\sqrt {{T_A}} {e^{ - i\ell\theta }} - \sqrt {{T_B}} {e^{i\ell\theta }}} \right|}^2}N} \right).
\label{9}
\end{equation}

\begin{figure}[htbp]
\centering
\includegraphics[width=8cm]{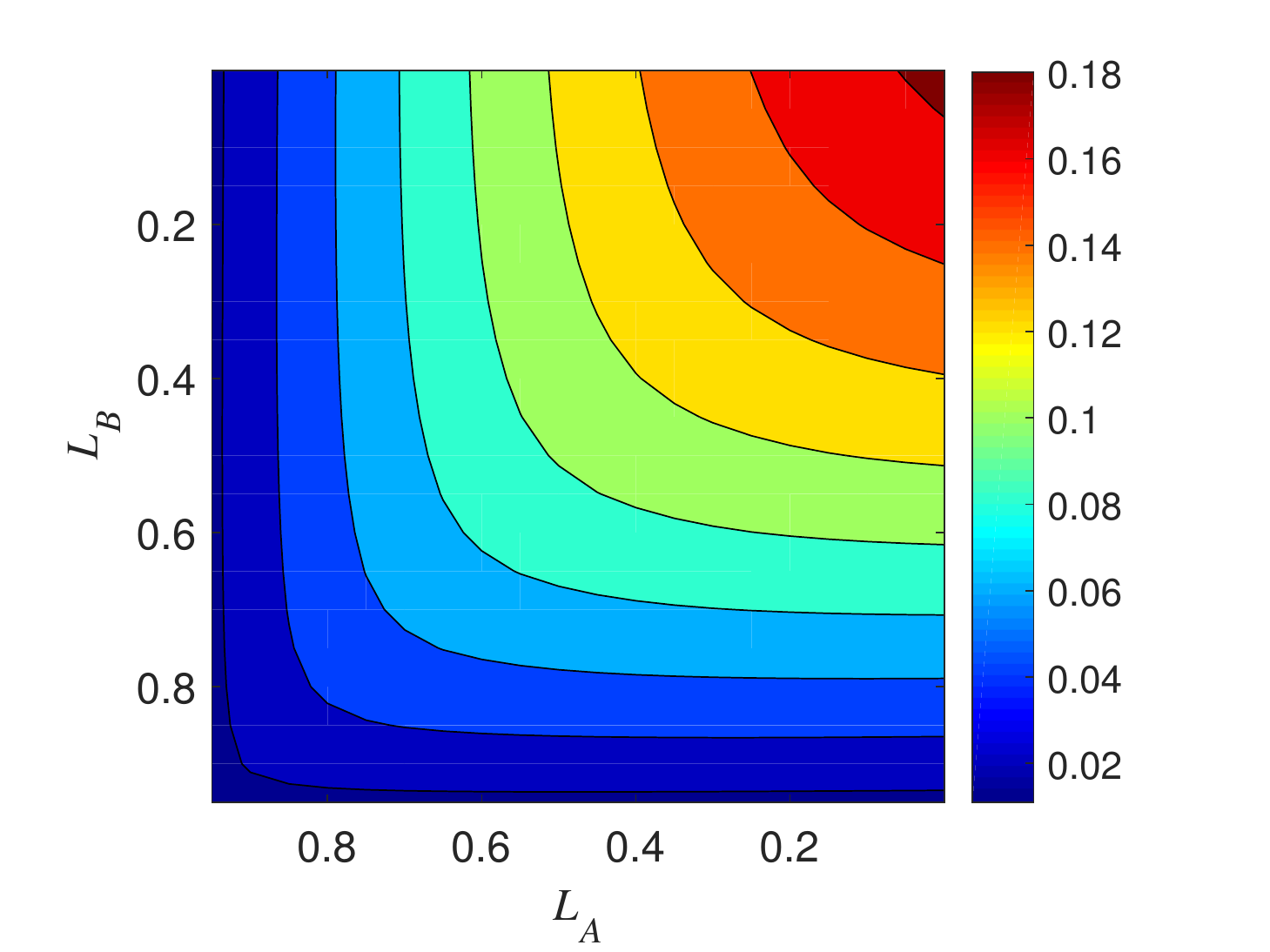}
\caption{The fidelity $H$ as a function of the transmission losses of two paths, $L_A$ and $L_B$. Where both $L_A$ and $L_B$ range from 0 to 1 and $N=3$.}
\label{f4}
\end{figure}

In terms of the probabilities of Eqs. (\ref{7}) and (\ref{9}), we describe the fidelity in the presence of the transmission loss, as manifested in Fig. \ref{f4}. 
A remarkable phenomenon is that, in the presence of identical total loss, the fidelity is better maintained when the two path losses are the same, i.e., the contour lines are convex compared with diagonal lines, $L_A+L_B={\rm constants}$. 
One can also find that the fidelity is reduced to 0 when either of the two paths are completely lossy. 
The reason for the above two phenomena is that the inconsistent losses erase the system's path indistinguishability$\---$the source of interference$\---$and then lead to the degeneration in the fidelity.

\subsection{Detection efficiency}

Next, we address the second scenario where the detection process is imperfect as well.
This scenario arises from photon loss in counting process, and it can be modeled by placing a beam splitter with transmissivity $\eta$ in front of the detector \cite{PhysRevA.83.063836}, where the parameter $\eta$ can be regarded as detection efficiency, and $1-\eta$ stands for the lossy ratio in the detection process.
Based on the state in Eq. (\ref{8}), the probability of zero photon output at port $B$ is written as
\begin{equation}
{p_{2}}\left( {\textrm{zero}{\rm{|}}\theta } \right) = \exp \left[ { - \eta {{\sin }^2}\left( {\ell\theta } \right)N} \right].
\label{10}
\end{equation}

\begin{figure}[htbp]
\centering
\includegraphics[width=8cm]{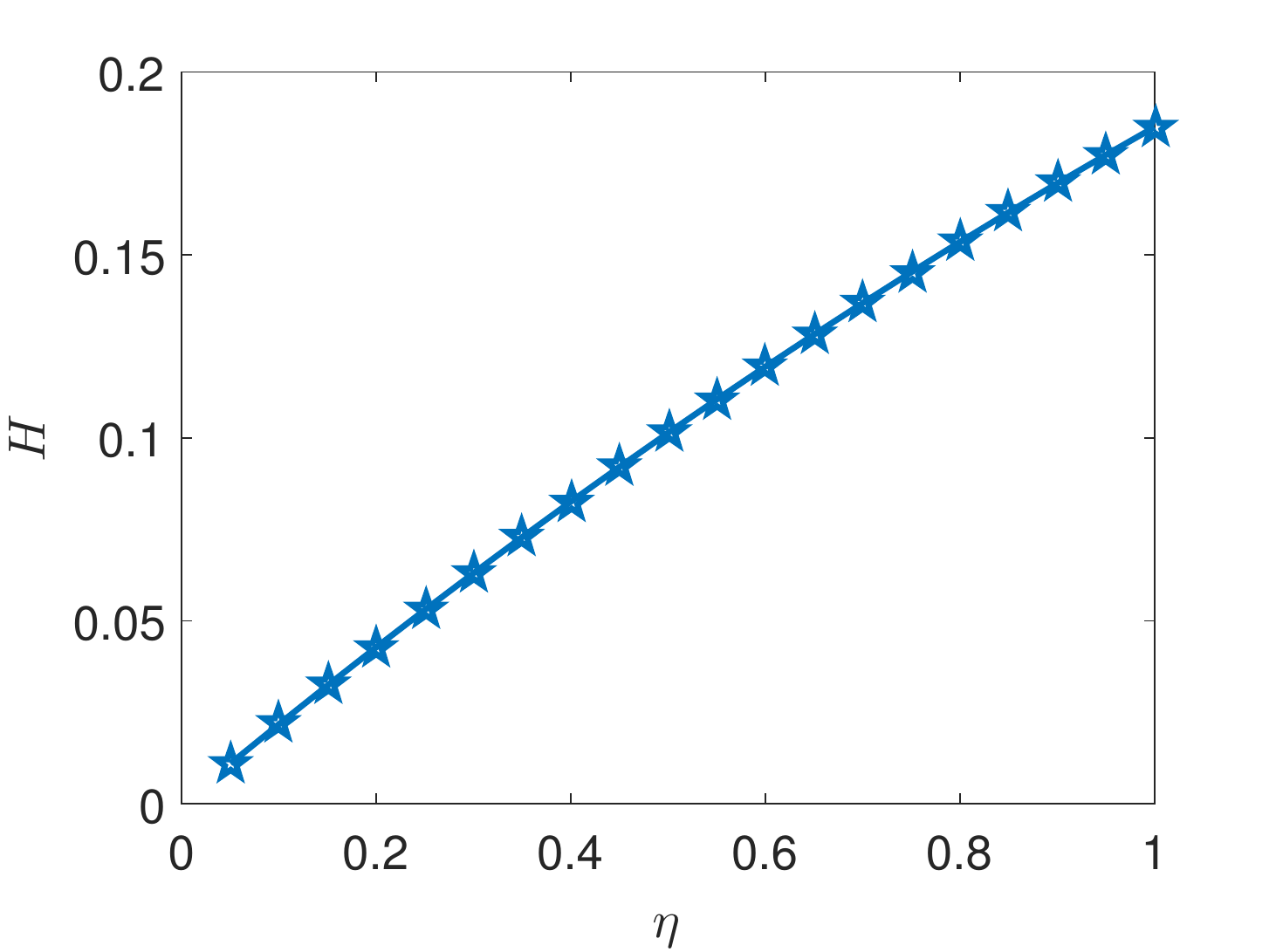}
\caption{The fidelity $H$ as a function of detection efficiency $\eta$, where $\eta$ ranges from $5\%$ to $100\%$ and $N=3$.}
\label{f5}
\end{figure}

In Fig. \ref{f5}, we plot the variation of the fidelity in the presence of imperfect detection efficiency.
It can be discovered that there is a better fidelity when the greater transmissivity exists. 
The relationship between the detection efficiency and the fidelity is approximately linear positive correlation. 
Notice that, with respect to the same losses in two paths ($T_A=T_B=T$), Eq. (\ref{9}) has an identical form with Eq. (\ref{10}) in the case of $T=\eta$.
This implies that detection loss and transmission loss are of a similar effect on the fidelity since either of them are linear loss.

\subsection{Dark counts}

Finally, another scenario about the realistic detectors, dark counts, is also discussed.
Assume that the rate of dark counts can be denoted as $r$. 
As a result, the probability of $n$ dark counts subjects to the following Poissonian distribution \cite{PhysRevA.95.053837}
\begin{equation}
{P_\textrm{dark}}\left( n \right) = {e^{ - r}}\frac{{{r^n}}}{{n!}}.
\label{11}
\end{equation}

The signal of Z detection is estimated from the photon number of the output outcome.
If the dark counts is zero, there is no change for detecting zero count.
Hence this probability arrives at
\begin{eqnarray}
\nonumber{p_{\rm{3}}}\left( {\textrm{zero}{\rm{|}}\theta } \right) &&= {p}\left( {\textrm{zero}{\rm{|}}\theta } \right){P_\textrm{dark}}\left( 0 \right) \\
&&= \exp \left[ { - {{\sin }^2}\left( {\ell\theta } \right)N-r} \right]
\label{12}
\end{eqnarray}

\begin{figure}[htbp]
\centering
\includegraphics[width=8cm]{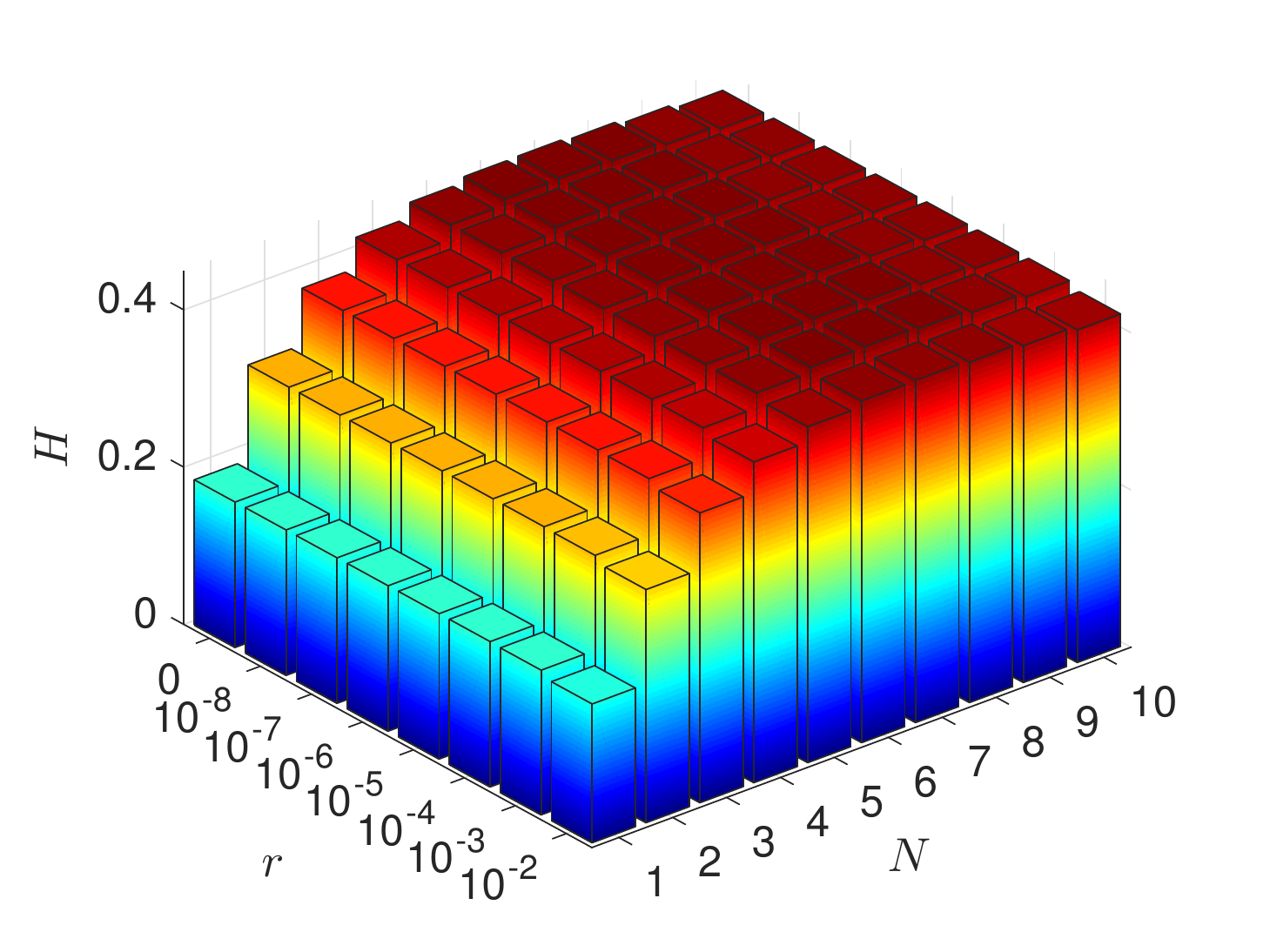}
\caption{The fidelity $H$ as a function of mean photon number $N$ and the rate of dark counts $r$, the range of $r$ is between ${\rm{1}}{{\rm{0}}^{ - 8}}$ and ${\rm{1}}{{\rm{0}}^{ - 2}}$, $N$ ranges from 1 to 10.}
\label{f6}
\end{figure}

Under the present technical conditions, where applicable, the range of $r$ is between ${\rm{1}}{{\rm{0}}^{ - 8}}$ and ${\rm{1}}{{\rm{0}}^{ - 2}}$ for a single APD.
In accordance with this range, we give the variation of the fidelity in the presence of the dark counts, as shown in Fig. \ref{f6}.
One discovers that the dark counts has almost no impact on the fidelity except for $r={\rm{1}}{{\rm{0}}^{ - 2}}$ (slight decrease).
Furthermore, the variation trend of the fidelity with mean photon number remains the same in spite of the existence of dark counts. 
Thus we can read that Z detection is a robust strategy for withstanding the disturbance of dark counts.

Based upon the consequents above, additionally, we briefly analyze the probability of zero count in a scenario that includes all the above realistic factors, hereon we let $T_A=T_B=T$.
In turn, this probability takes the form:
\begin{equation}
{p^{*}}\left( {\textrm{zero}{\rm{|}}\theta } \right) = \exp \left[ { - {{\sin }^2}\left( {\ell\theta } \right){N_{\rm e}}-r} \right],
\label{13}
\end{equation}
where ${N_{\rm e}}=T\eta N$ denotes effective mean photons triggering the Gm-APD. 
This implies that we can utilize ${N_{\rm e}}$ to perform a precise estimation, whereas exact input photon number $N$ is dispensable, and it is generally difficult to precisely assess the photon loss in an actual estimation process.
In addition, the probability in Eq. (\ref{13}) cannot reach its maximum because of the term $e^{-r}$, if and only if the minimum approaches to 0, the visibility can approximatively stay the same.

\section{Experimental implementation based on Bayesian estimation}
\label{s5}

As the last part of the work in this paper, we perform a proof of principle experiment for Z detection and adopt Bayesian estimation \cite{APD}. 
For this strategy, the conditional probability$\---$a posteriori distribution$\---$of the measuring sample is the core in our estimation protocol.
It is easy to obtain from the Bayesian theorem, which states that  $p\left( {\theta {\rm{|}}W} \right)p\left( W \right) = p\left( {W{\rm{|}}\theta } \right)p\left( \theta  \right)$, where $p\left( \theta  \right)$ is the priori distribution which has defined in Eq. (\ref{6}), and $p\left( W \right)$ is the overall probability of the measuring sample. 
Therefore, we can write this posteriori distribution as 
\begin{equation}
p\left( {\theta {\rm{|}}W} \right) = \frac{1}{G}\mathop \prod \limits_{k = 1}^M p\left( {{W_k}{\rm{|}}\theta } \right),
\label{14}
\end{equation}
where $G$ is responsible for the normalization, it can be calculated by
\begin{equation}
G = \int_{ - \pi }^\pi  {\mathop \prod \limits_{k = 1}^M p\left( {{W_k}{\rm{|}}\theta } \right)} {\kern 1pt} d\theta
\label{15}
\end{equation}

One knows that Bayesian estimator is asymptotically unbiased. 
For $M \gg 1$, the Eq. (\ref{14}) can be rewritten as
\begin{eqnarray}
\nonumber  p\left( {\theta {\rm{|}}{\theta ^ * }} \right) =&& \frac{1}{G}\exp \left[ {M {p\left( {\textrm{nonzero}{\rm{|}}{\theta ^ * }} \right)\log p\left( {\textrm{nonzero}{\rm{|}}\theta } \right) } } \right]\\
&&\times\exp \left[ {M{p\left( {\textrm{zero}{\rm{|}}{\theta ^ * }} \right)\log p\left( {\textrm{zero}{\rm{|}}\theta } \right) } } \right],
\label{16}
\end{eqnarray}
where $\theta ^ * $ represents the actual value of the angular displacement.

\begin{figure}[htbp]
	\centering
	\includegraphics[width=8cm]{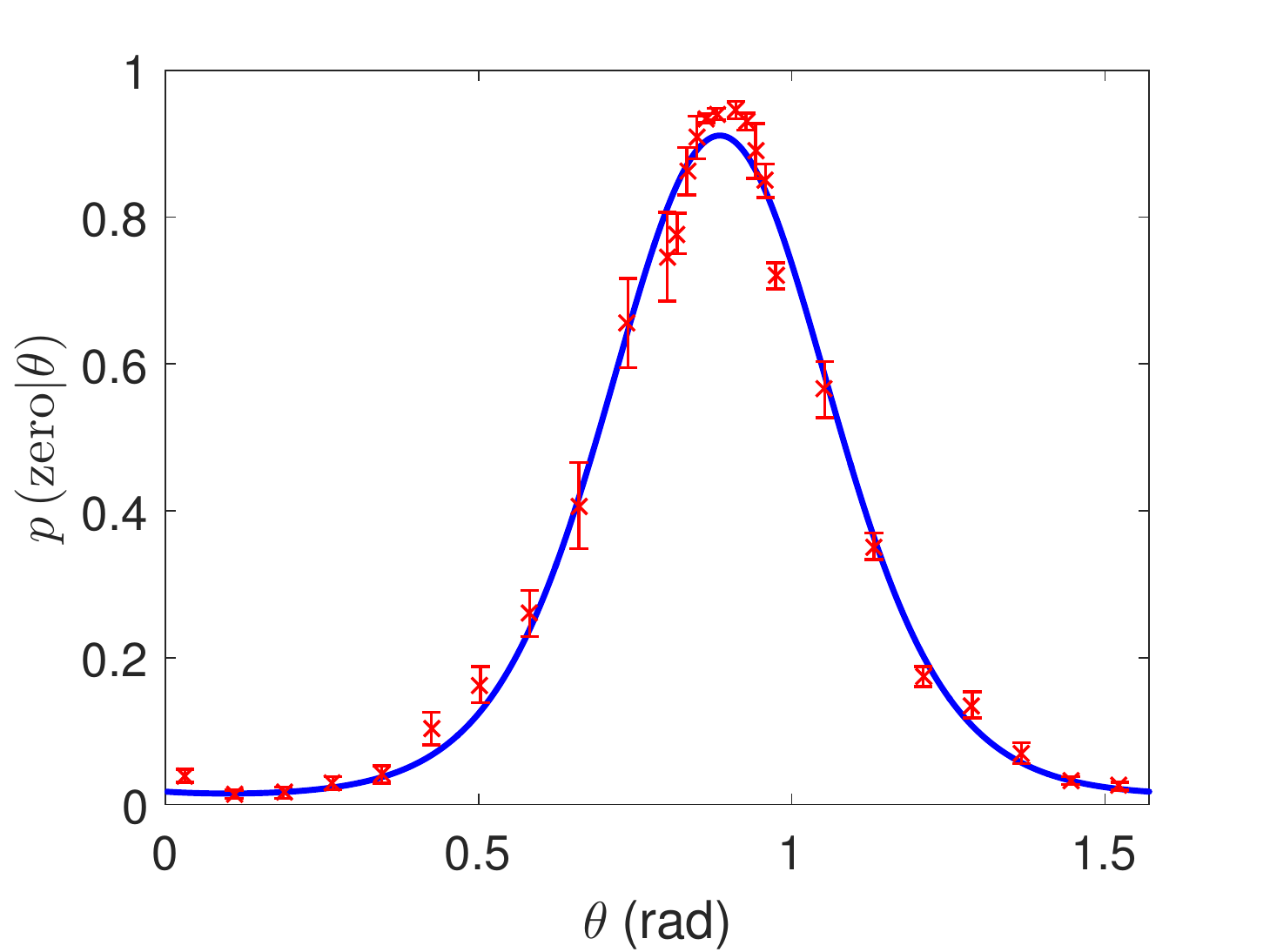}
	\caption{Probability of zero count (normalized measuring counts) as a function of angular displacement. The solid red dots are experimental measuring data, while the solid blue line is a fit toward the data. Error bars are one standard deviation due to propagated Poissonian statistics, and each one is obtained by 20 trials with $M=2000$ in every trial.}
	\label{f7}
\end{figure}

By rotating the Dove prism, we record the number of zero count detected by the Gm-APD at several points, where $\ell=2$ is used in experiment.
The mean and standard deviation for each sampling point are also provided.
Then, based upon least-square method, we fit a signal curve for Z detection, as displayed in Fig. \ref{f7}, detailed expression is
\begin{eqnarray} 
{p}\left( {\textrm{zero}|\theta } \right) =0.911\exp \left\{ { - 4.11{{\sin }^2}\left[ {2\left( \theta+0.686\right)  } \right]} \right\}
\label{17}
\end{eqnarray}
with $N_{\rm e}=4.11$.  
According to the definition of visibility \cite{dowling2008quantum}, one finds that the visibility of experimental outcome is 96.7\%.
Furthermore, the full wave at half maximum is 0.4224 radians, which is a super-resolved signal enhancement by a factor of 1.86 compared to the Rayleigh limit.

\begin{figure}[htbp]
	\centering
	\includegraphics[width=8cm]{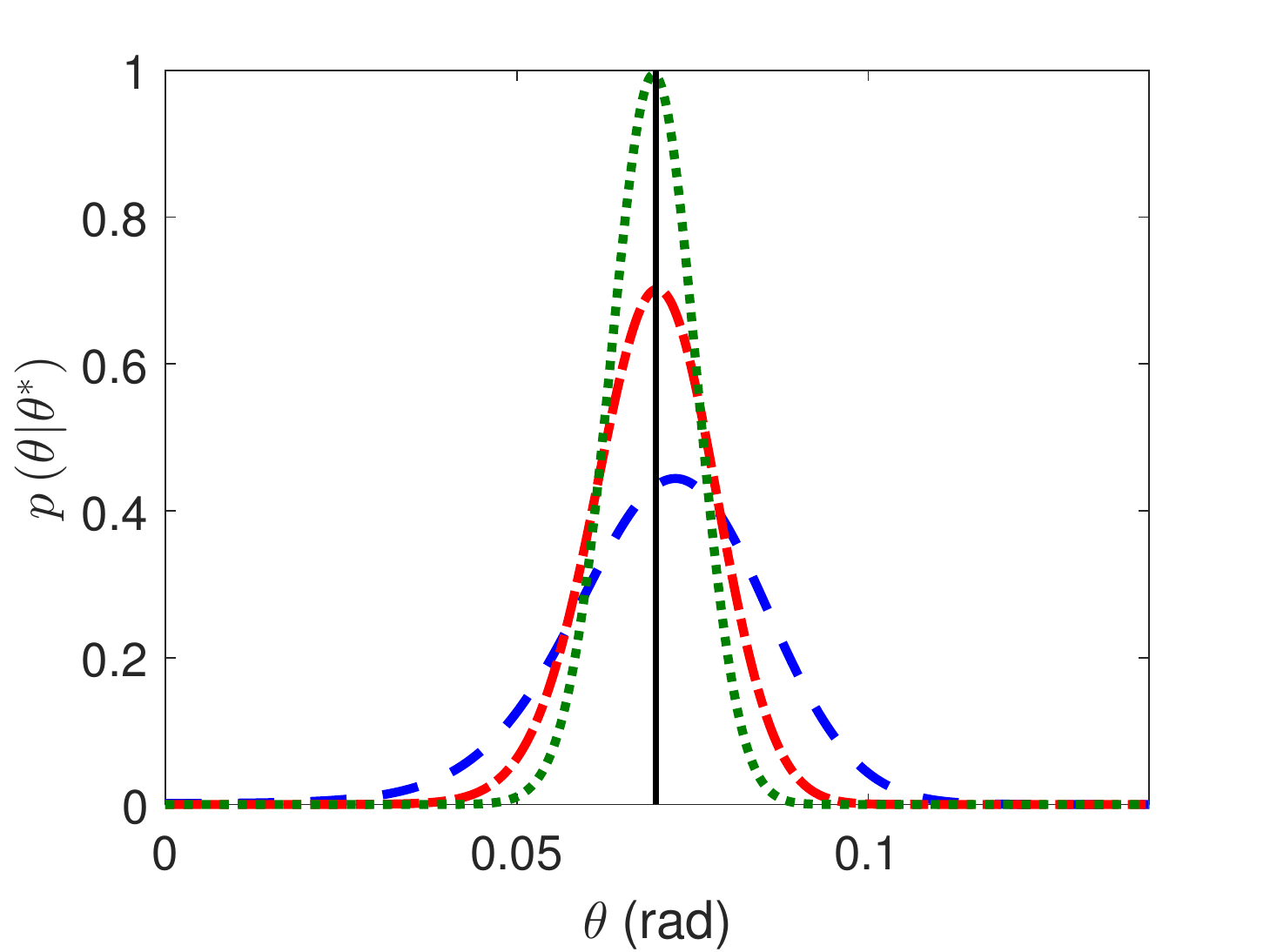}
	\caption{The posteriori distribution $p\left( {\theta {\rm{|}}{\theta ^ * }} \right)$ as a function of angular displacement $\theta$, where actual value $\theta^*$ sits at 0.0698. 
		The blue dashed line consists of $M = 200$ Z detection, of which 167 are zero.	
		$M = 500$ and $M = 1000$ correspond to red and green dashed lines, respectively, zero outcomes are 419 and 841.}
	\label{f8}
\end{figure}

However, the maximum of the signal is less than 1, this means that interference is not thorough, i.e., incomplete extinction.
Imitating the model in Ref. \cite{Istrati2014Super} and Eq. (\ref{13}), Eq. (\ref{17}) can be written as
\begin{eqnarray} 
{p}\left( {\textrm{zero}|\theta } \right) = \exp \left\{ { - 4.11{{\sin }^2}\left[ {2\left( \theta+0.686\right)  } \right]-{n_{\rm b}}} \right\},
\label{18}
\end{eqnarray}
where $n_{\rm b}=0.093$, corresponding to $\exp\left(-{n_{\rm b}}\right) =0.911 $, is background noise, including dark counts and the photons without participating in interference.

The main reasons for this phenomenon are as follows: initially, the conversion efficiency of the spiral phase plate for the beam is less than 100\%, and non-OAM beam is not modulated by Dove prism.
Furthermore, for the polarizing beam splitters, the extinction ratio is also a non-ideal value which is less than 100\%.
Eventually, the linearly polarized state will be changed slightly due to the rotation of Dove prism, it introduces a component which is perpendicular to the input polarization.

For an actual value $\theta ^ *$, in experiment we perform repeated trials $M$ with 200, 500 and 1000, respectively.
Combined with the conditional probability formula of Z detection in experiment and Eq. (\ref{15}), in Fig. \ref{f8} we provide the posteriori distribution for our estimation.
Figure \ref{f8} indicates that the posteriori distribution becomes obvious for actual value with the increase of trials, i.e., the probabilities of other values gradually tend to 0. 
For a large enough $M$, the posteriori distribution can be approximately evolved into $p\left( {\theta {\rm{|}}{\theta ^ * }} \right) = \delta \left(\theta ^ *\right)$.
In addition, a classical effect is that the sensitivity is enhanced by increasing trials. Meanwhile, from Fig. \ref{f8} we can find that it will draw the estimated value closer to the actual one with large trials.

\section{Conclusion}
\label{s6}

In summary, we study the optimal binary strategy toward angular displacement estimation based on fidelity appraisal, which is a method that can replace standard deviation metric.
Two binary detection strategies$\---$Z detection and parity detection$\---$are analyzed, and the fidelity of Z detection has an overwhelming advantage compared to that of parity detection.
On the basis of the optimal detection strategy, we additionally discuss the effects of several realistic scenarios on fidelity, containing transmission loss, detection efficiency, dark counts, and those which are a combination thereof.	
For the transmission loss, identical losses in the two paths gives a better fidelity than inconsistent losses in the case of uniform total loss.
An approximately positive correlated relationship between the detection efficiency and the fidelity is proved by analysis.
We also find that the effect of dark counts on the fidelity is almost negligible. 
Finally, the proof of principle experiment is carried out, and we perform Bayesian estimation on measuring data.
A boosted super-resolving signal with a factor of 1.86 is shown, and the actual angular displacement can be precisely estimated with increasing the number of trials $M$.
Overall, the experimental results are in good agreement with theoretical analysis.

\section*{Acknowledgments}
This work is supported by the National Natural Science Foundation of China (Grant No. 61701139).

%


\begin{thebibliography}{39}
\bibitem [{APD()}]{APD}%
	\BibitemOpen
	\href@noop {} {\bibinfo  {journal} {The type of the Gm-APD in experiment is
			COUNT-100C-FC. The rate of dark count is 20c/s and photon detection
			efficiency Pd @ 670 nm is 84\%. In addition, afterpulsing is 0.1\% and dead
			time is 43 ns}\ }\BibitemShut {NoStop}%
\end{thebibliography}

\end{document}